\documentclass[fleqn,twoside,twocolumn]{revtex4} % Specifies the document class %,unsortedaddress
 \usepackage{amsmath,amssymb}
\numberwithin{equation}{section}
\begin{document}
\title[THE LAGRANGIAN AND HAMILTONIAN FORMALISMS]%колонтитул
{THE LAGRANGIAN AND HAMILTONIAN FORMALISMS FOR THE CLASSICAL
RELATIVISTIC ELECTRODYNAMICS MODELS\\ REVISITED
}%
\author{N.N. Bogolubov (jr.),}%1 автор
\affiliation{V.A. Steklov Mathematical Institute of RAN}%институт
\address{Moscow, Russian
Federation}%адрес
\affiliation{Abdus Salam International Centre for Theoretical Physics}%институт
\address{Trieste, Italy}%адрес
\email{nikolai_bogolubov@hotmail.com}%e-mail
\author{A.K. Prykarpatsky}%
\affiliation{AGH University of Science and Technology}%
\address{Krak\'{o}w 30-059, Poland}%
\affiliation{Ivan Franko State Pedagogical
University}%
\address{Drohobych, Lviv region, Ukraine}
\email{pryk.anat@ua.fm, prykanat@cybergal.com}%e-mail%
%\avtorcol{N.N. BOGOLUBOV (jr.), A.K. PRYKARPATSKY}%%колонтитул
%\udk{533.9} \pacs{11.10.Ef} \razd{\seci}
\begin{abstract}
The work is devoted to studying some new classical electrodynamics
models of interacting charged point particles and the aspects
of the quantization via the Dirac procedure related to them. Based on the
vacuum field theory no-geometry approach developed in [6,7,9],
the Lagrangian and Hamiltonian reformulations of some
alternative classical electrodynamics models are devised. The
Dirac-type quantization procedure for the considered alternative
electrodynamics models, based on the obtained canonical
Hamiltonian formulations, is developed.
\end{abstract}

\setcounter{page}{753}%

\maketitle

\section{Introductory Setting}

It is well known [5, 11, 13, 42] that the ``\textit{quintessence}''
of
the classical relativistic electrodynamics in the Minkowski space $\mathbb{M}%
^{4}:=\mathbb{E}^{3}\mathbb{\times R}$ of a moving charged point particle
consists in a successive derivation of the Lorentz force expression%
%1.1
\begin{equation}
F:=qE+qu\times B,  \label{0.1}
\end{equation}
where $q\in \mathbb{R}$ is a particle electric charge, $u\in \mathbb{E}^{3}$
is its velocity vector (expressed here in the light speed $c$ units),%
%1.2
\begin{equation}
E:=-\partial A/\partial t-\nabla \varphi  \label{0.2}
\end{equation}
is the corresponding external electric field and
%1.3
\begin{equation}
B:=\nabla \times A  \label{0.3}
\end{equation}
is the corresponding external magnetic field acting on the charged
particle, being expressed through the suitable vector $A:\mathbb{M}%
^{4}\rightarrow \mathbb{E}^{3}$ and scalar $\varphi :\mathbb{M}%
^{4}\rightarrow \mathbb{R}$ potentials. Writing fields (\ref{0.2}) and (\ref%
{0.3}), we have denoted, by ``$\nabla $,'' the standard gradient
operation with respect to the spatial variable $r\in \mathbb{E}^{3}$
and, by ``$\times $,'' the vector product in the three-dimensional
Euclidean vector space $\mathbb{E}^{3}$  naturally endowed with the
usual scalar product $\langle\cdot ,\cdot \rangle.$

Let the additional Lorentz condition
%1.4
\begin{equation}
\partial \varphi /\partial t+\langle\nabla ,A\rangle=0  \label{0.4}
\end{equation}
for the potentials $(\varphi ,A):\mathbb{M}^{4}\rightarrow \mathbb{R\times E}%
^{3}$ satisfying the Lorentz-invariant wave field equations
%1.5
\begin{equation}
\partial ^{2}\varphi /\partial t^{2}-\nabla ^{2}\varphi =\rho ,
\partial ^{2}A/\partial t^{2}-\nabla ^{2}A=J  \label{0.5}
\end{equation}
be imposed. Here, $\rho :\mathbb{M}^{4}\rightarrow \mathbb{R}$ and $J:\mathbb{M}%
^{4}\rightarrow \mathbb{E}^{3}$ are, respectively, charge and current
densities of the ambient matter satisfying the charge continuity
relation%
%1.6
\begin{equation}
\partial \rho /\partial t+\langle\nabla ,J\rangle=0.  \label{0.7}
\end{equation}
Then the well-known [13, 5, 42, 11] classical electromagnetic
Maxwell field equations
%1.7
\[\nabla \times E+\partial B/\partial t =0, \quad \langle\nabla ,E\rangle=\rho,\]
\begin{equation}
\nabla \times B-\partial E/\partial t =J,\quad \nabla \times B=0
\label{0.6}
\end{equation}
hold for all $(r,t)\in \mathbb{M}^{4}$ with respect to a chosen reference
system $\mathcal{K}.$

We note that the Maxwell equations (\ref{0.6}) are not directly reduced, via
definitions (\ref{0.2}) and (\ref{0.3}), to the wave field equations (\ref%
{0.5}), if the Lorentz condition (\ref{0.4}) is not taken into
consideration. This fact becomes very important if we change our subject of
reasonings to the determining role of the Maxwell equations (\ref{0.6})
with (\ref{0.7}) and put the
Lorentz condition (\ref{0.4}) jointly with (\ref{0.5}) and the continuity
relation (\ref{0.7}) into consideration as governing relations.
Concerning the assumptions formulated above, the
following proposition holds.

\noindent{\bf Proposition 1.}{\it The Lorentz-invariant wave
equations (\ref{0.5}) for the potentials $(\varphi
,A):\mathbb{M}^{4}\rightarrow \mathbb{R\times E}^{3}$ considered
jointly
with the Lorentz condition (\ref{0.4}) and the charge continuity relation (%
\ref{0.5}) are completely equivalent to the Maxwell field equations (\ref%
{0.6}).}

\noindent{\sc Proof.} Really, having substituted the partial
derivative $\partial \varphi /\partial t=-\langle\nabla ,A\rangle$
following from (\ref{0.4}) into (\ref{0.5}), one easily obtains that
%1.8
\begin{equation}
\partial ^{2}\varphi /\partial t^{2}=-\langle\nabla ,\partial A/\partial
t\rangle=\langle\nabla ,\nabla \varphi \rangle+\rho .  \label{0.8}
\end{equation}%
Relation (\ref{0.8}) yields the gradient expression
%1.9
\begin{equation}
\langle\nabla ,-\partial A/\partial t-\nabla \varphi \rangle=\rho
. \label{0.9}
\end{equation}
Taking the electric field definition (\ref{0.2}) into account, expression (%
\ref{0.9}) is reduced to
%1.10
\begin{equation}
\langle\nabla ,E\rangle=\rho ,  \label{0.10}
\end{equation}
entering the first pair of the Maxwell equations (\ref{0.6}).

Having now applied the operation ``$\nabla \times $'' to (\ref{0.2}%
), we obtain, owing definition (\ref{0.3}),
%1.11
\begin{equation}
\nabla \times E+\partial B/\partial t=0,  \label{0.11}
\end{equation}
being the next equation entering the first pair of the Maxwell equations (%
\ref{0.6}).

Applying now the operation ``$\nabla \times $'' to (\ref{0.3}):%
%1.12
\[\nabla \times B =\nabla \times (\nabla \times A)=\nabla
\langle\nabla ,A\rangle-\nabla ^{2}A=\]
\[=-\nabla (\partial \varphi /\partial t)-\partial ^{2}A/\partial
t^{2}+(\partial ^{2}A/\partial t^{2}-\nabla ^{2}A)=\]
\begin{equation}
=\frac{\partial }{\partial t}(-\partial \varphi /\partial
t-\partial A/\partial t)+J=\partial E/\partial t+J,  \label{0.12}
\end{equation}
we now obtain the resulting equation
\[\nabla \times B=\partial E/\partial t+J\]
exactly coinciding with that entering the second pair of the
Maxwell equations (\ref{0.6}). The final \textit{``no-magnetic
charge}'' equation
\[\langle\nabla ,B\rangle=\langle\nabla ,\nabla \times
A\rangle=0\] entering (\ref{0.6}) follows easily from the elementary
differential identity $\langle\nabla ,\nabla \times \rangle=0,$
thereby finishing the proof.

\hspace*{7.9cm}$\square$

The proposition above allows us to consider the potential functions $(\varphi
,A):\mathbb{M}^{4}\rightarrow \mathbb{R\times E}^{3}$ as fundamental
ingredients of the ambient vacuum field medium, by means of which we can try
to describe the related physical behavior of charged point particles
imbedded into the space-time $\mathbb{M}^{4}$ filled in with the vacuum field
medium. This way of reasoning is strongly supported by the next important
observation.

\textbf{Observation.} \textit{The Lorentz condition (\ref{0.4}) means,
in reality, the scalar potential field }$\varphi :$\textit{\
}$\mathbb{M} ^{4}\rightarrow \mathbb{R}$\textit{\ continuity
relation, whose origin lies in some new field conservation
law characterizing the deep intrinsic vacuum field medium
structure.}

To make this observation more clear and exact, let us recall the definition
[13,42,5,11] of the electric current $J:$\textit{\ }$\mathbb{M}%
^{4}\rightarrow \mathbb{E}^{3}$ in the dynamical form
%1.13
\begin{equation}
J:=\rho v,  \label{0.13}
\end{equation}
where the vector $v:$ \textit{\ }$\mathbb{M}^{4}\rightarrow
\mathbb{E}^{3}$ is the corresponding electric charge flow
velocity understood  here [18]
in the hydrodynamical sense. Thus, the continuity relation%
%1.14
\begin{equation}
\partial \rho /\partial t+\langle\nabla ,\rho v\rangle =0  \label{0.14}
\end{equation}
holds, and it can be easily rewritten \cite{MC} \ as the integral
conservation law
%1.15
\begin{equation}
\frac{d}{dt}\int\limits_{\Omega _{t}}\rho d^{3}r=0  \label{0.15}
\end{equation}
for the charge held inside of any bounded domain $\Omega _{t}\subset \mathbb{%
E}^{3}$  moving in the space-time $\mathbb{M}^{4}$. Its every
intrinsic point $r\in \Omega _{t}$ changes in time  as
%1.16
\begin{equation}
dr/dt:=v  \label{0.16}
\end{equation}
with respect to the corresponding electric charge flow velocity.

Following the same reasonings as above, we can state the next
proposition.

\noindent{\bf Proposition 2.} {\it The Lorentz condition
(\ref{0.4}) is exactly equivalent to the following integral
conservation law:}
%1.17
\begin{equation}
\frac{d}{dt}\int\limits_{\Omega _{t}}\varphi d^{3}r=0,  \label{0.17}
\end{equation}
{\it where $\Omega _{t}\subset \mathbb{R}^{3}$  is any bounded
domain moving according to the evolution equation}
%1.18
\begin{equation}
dr/dt:=v,  \label{0.18}
\end{equation}
{\it meaning the velocity vector of a local potential field flow
inside the domain $\Omega _{t}$ propagating in the space-time
$\mathbb{M}^{4}$.}

\noindent{\sc Proof.} Consider, first, the corresponding solutions
to the potential field equations (\ref{0.5}), taking condition
(\ref{0.13}) into account. Owing to the results from [5, 13], one
finds that
%1.19
\begin{equation}
A=\varphi v,  \label{0.19}
\end{equation}
giving rise to the following form of the Lorentz condition (\ref{0.4}):%
%1.20
\begin{equation}
\partial \varphi /\partial t+\langle\nabla ,\varphi v\rangle=0.  \label{0.20}
\end{equation}
The latter, obviously, can be equivalently rewritten [18] as the
integral conservation law (\ref{0.17}).

\hspace*{7.9cm}$\square$\vskip3mm

The proposition formulated above allows a physically motivated
interpretation by means of involving the important notion -
\textit{the vacuum potential field} describing the observable
interactions between charged point particles. Namely, we can
\textit{a priori} endow the ambient vacuum medium with a scalar
potential field function $W:=q\varphi : \mathbb{M}^{4}\rightarrow
\mathbb{R}$ satisfying the governing vacuum field equations
%1.21
\begin{equation}
\partial ^{2}W/\partial t^{2}-\nabla ^{2}W=0,\text{ \ \ }\partial W/\partial
t+\langle\nabla ,Wv\rangle=0,  \label{0.21}
\end{equation}
taking into account that there are no external sources besides
material particles possessing only a virtual capability to disturb
the vacuum field medium. Moreover, as the related vacuum potential
field function $W:\mathbb{M }^{4}\rightarrow \mathbb{R}$ allows the
natural potential energy interpretation, its origin should be
assigned not only to the charged interacting medium, but also to any
other medium possessing a virtual capability to interact, including,
for instance, material particles interacting through the gravity.

The latter allows us to make the next important step consisting in
deriving the equation governing the corresponding potential field
$\bar{W}:\mathbb{M} ^{4}\rightarrow \mathbb{R}$ assigned to the
vacuum field medium in the vicinity of any particle located at a
point $R(t)\in \mathbb{E}^{3}$ moving with velocity $u\in
\mathbb{E}^{3}$ at time $t\in \mathbb{R}.$ As it can be easily
enough shown [6, 7, 10], the corresponding evolution equation
governing the related potential field function $\bar{W}:\mathbb{M}
^{4}\rightarrow \mathbb{R}$ looks as follows:
%1.22
\begin{equation}
\frac{d}{dt}(-\bar{W}u)=-\nabla \bar{W},  \label{0.22}
\end{equation}
where, by definition, $\bar{W}:=W(r,t)|_{r\rightarrow R(t)},$ $u:=dR(t)/dt$
at a point particle location $(R(t),t)\in \mathbb{M}^{4}.$

Similarly, if we consider two point particles which interact with each other, are
located at points $R(t)$ and $R_{f}(t)\in
\mathbb{E}^{3}$ at time $t\in \mathbb{R}$, and are moving,
respectively, with velocities $u:=dR(t)/dt$ \ and \
$u_{f}:=dR_{f}(t)/dt,$ the corresponding potential field function
$ \bar{W}:\mathbb{M}^{4}\rightarrow \mathbb{R}$ related to the
particle located at the point $R(t)\in \mathbb{E}^{3},$ looks as
%1.23
\begin{equation}
\frac{d}{dt}[-\bar{W}(u-u_{f})]=-\nabla \bar{W}.  \label{0.23}
\end{equation}
The dynamical potential field equations (\ref{0.22}) and (\ref{0.23}) obtained
above stimulate us to proceed the further study of their physical properties
and to compare them with the already available classical results for suitable Lorentz-type
forces described within the electrodynamics of moving charged point
particles interacting with an external electromagnetic field.

To realize this program, we, being strongly inspired by works
[22--26, 33, 40] and especially by the original works [28, 29]
devoted to solving the classical problem of reconciling
gravitational and electrodynamical charges within the Mach--Einstein
ether paradigm, first reanalyze successively the classical
Mach--Einstein relativistic electrodynamics of a moving charged
point particle and, second, study the resulting electrodynamic
theories associated with our vacuum potential field dynamical
equations (\ref{0.22}) and (\ref{0.23}), making use of the
fundamental Lagrangian and Hamiltonian formalisms. Based on the
results obtained, the canonical Dirac-type quantization procedure is
applied to the corresponding energy conservation laws related
naturally to electrodynamic models considered in the work.

\section{Classical Relativistic Electrodynamics Revisited}

The classical relativistic electrodynamics of a freely moving
charged point particle in the Minkowski space-time
$\mathbb{M}^{4}:=\mathbb{E}^{3}\times \mathbb{R}$ is, as well known,
based [5, 11, 13, 42] \ on the Lagrangian formalism assigning the
Lagrangian function
%2.1
\begin{equation}
\mathcal{L}:=-m_{0}(1-u^{2})^{1/2}  \label{1.1}
\end{equation}
to it, where $m_{0}\in \mathbb{R}$ is the so-called particle rest
mass, and $u\in \mathbb{E}^{3}$ is its spatial velocity in the
Euclidean space $\mathbb{E} ^{3}$ expressed here and throughout
further in the light speed $c$ units. The least action Fermat
principle in the form
%2.2
\begin{equation}
\delta S=0,\text{ \ \
}S:=-\int\limits_{t_{1}}^{t_{2}}m_{0}(1-u^{2})^{1/2}dt \label{1.2}
\end{equation}
for any fixed temporal interval $[t_{1},t_{2}]\subset \mathbb{R}$ gives
rise to the well-known relativistic relations for the
particle mass
%2.3
\begin{equation}
m=m_{0}(1-u^{2})^{-1/2},  \label{1.3}
\end{equation}
the particle momentum
%2.4
\begin{equation}
p:=mu=m_{0}u(1-u^{2})^{-1/2}, \label{1.4}
\end{equation}
and the particle energy
%2.5
\begin{equation}
\mathcal{E}_{0}=m=m_{0}(1-u^{2})^{-1/2}.  \label{1.5}
\end{equation}
The origin of Lagrangian (\ref{1.1}) can be extracted, owing to the reasonings from [13,42],
from the action expression%
%2.6
\begin{equation}
S:=-\underset{t_{1}}{\overset{t_{2}}{\int
}}m_{0}(1-u^{2})^{1/2}dt=-\underset {\tau _{1}}{\overset{\tau
_{2}}{\int }}m_{0}d\tau ,  \label{1.6}
\end{equation}
on the suitable temporal interval $[\tau _{1,}\tau _{2}]\subset
\mathbb{R} \mathbf{,}$ where,\textbf{\ }by definition,
%2.7
\begin{equation}
d\tau :=dt(1-u^{2})^{1/2}  \label{1.6a}
\end{equation}
and $\tau \in \mathbb{R}$ is the so-called proper temporal parameter
assigned to a freely moving particle with respect to the  ``rest''
reference system $ \mathcal{K}_{r}.$  Action (\ref{1.6}) looks from
the dynamical point of view slightly controversial, since it is
physically defined with respect to the  ``rest''  reference system
$\mathcal{K}_{r},$ giving rise to the constant action $S=-m_{0}(\tau
_{2}-\tau _{1}),$ as the limits of integrations $ \tau _{1}<\tau
_{2}\in \mathbb{R}$ were taken to be fixed from the very beginning.
Moreover, let us consider this particle as charged with a charge $
q\in \mathbb{R}$ and moving in the Minkowski space-time
$\mathbb{M}^{4}$ under the action of an electromagnetic field
$(\varphi ,A)\in \mathbb{R}\times \mathbb{E}^{3},$ and let the
corresponding classical (relativistic) action functional be chosen\
(see [5, 11, 13, 26, 42])  as
%2.8
\begin{equation}
S:=\underset{\tau _{1}}{\overset{\tau _{2}}{\int }}[-m_{0}d\tau +q\langle A,\dot{r}%
\rangle d\tau -q\varphi (1-u^{2})^{-1/2}d\tau ]  \label{1.7}
\end{equation}
with respect to the so-called \ ``rest'' reference system
parametrized by the Euclidean space-time variables $(r,\tau )\in
\mathbb{E}^{4}$. Here, as before, $\langle\cdot ,\cdot \rangle$ is
the standard scalar product in the related Euclidean subspace
$\mathbb{E}^{3}$, and we denoted $\dot{r}:=dr/d\tau $ in contrast to
the definition $u:=dr/dt.$ Action (\ref{1.7}) can be rewritten with
respect to the reference system moving with a velocity $u\in
\mathbb{E} ^{3}$ as
%2.9
\begin{equation}
S=\underset{t_{1}}{\overset{t_{2}}{\int }}\mathcal{L}dt,\text{ \
}\mathcal{L} :=-m_{0}(1-u^{2})^{1/2}+q\langle A,u\rangle -q\varphi
 \label{1.8}
\end{equation}
on the suitable temporal interval  $[ t_{1},t_{2}]\subset
\mathbb{R}$. This gives rise to the  [13, 5, 42, 11] dynamical
expressions
%2.10
\begin{equation}
P=p+qA,\text{ \ \ \ \ }p=mu,  \label{1.9}
\end{equation}
for the particle momentum and
%2.11
\begin{equation}
\mathcal{E}_{0}=[m_{0}^{2}+(P-qA)^{2}]^{1/2}+q\varphi  \label{1.10}
\end{equation}
for the particle energy. Here, by definition, $P\in \mathbb{E}^{3}$ means
the common momentum of the particle and the ambient electromagnetic field at
a space-time point $(r,t)\in \mathbb{M}^{4}.$

The obtained expression (\ref{1.10}) for the particle energy
$\mathcal{E} _{0} $ also looks slightly controversial, since the
potential energy $ q\varphi ,$ entering additively, has no impact to
the particle mass $ m=m_{0}(1-u^{2})^{-1/2}.$ As it was already
mentioned \cite{Br} by L. Brillouin, the fact that the potential
energy has no impact to the particle mass says us that ``... any
possibility of existing the particle mass related to an external
potential energy is completely excluded''. This and some other
special relativity theory and electrodynamics problems, as is well
known, stimulated many other prominent physicists of the past [4,
16, 21, 43, 42] and the present [22--24, 33, 38--41, 44, 45] to make
significant efforts aiming to develop alternative relativity
theories based on completely different [20,  25, 28, 30--32, 34, 35,
48, 49] space-time and matter structure principles.

There also is another controversial inference from the action
expression (\ref{1.8}). As one can easily show [5, 11, 13, 42], the
corresponding dynamical equation for the Lorentz force is given as
%2.12
\begin{equation}
dp/dt=F:=qE+qu\times B,  \label{1.11}
\end{equation}
where the operation $``\times ''$ denotes, as before, the standard
vector product, and we put, by definition,
%2.13
\begin{equation}
E:=-\partial A/\partial t-\nabla \varphi  \label{1.12}
\end{equation}
for the related electric field and
%2.14
\begin{equation}
B:=\nabla \times A  \label{1.13}
\end{equation}
for the related magnetic field acting on the charged point particle
$q;$ the operation ``$\nabla $'' is here, as before, the standard
gradient. The obtained expression (\ref{1.11}) means, in particular,
that the Lorentz force $F$\ depends linearly on the particle
velocity vector $u\in \mathbb{E} ^{3},$ giving rise to its strong
dependence on the reference system, with respect to which the
charged particle $q$ moves. Namely, the attempts to reconcile this
and some related controversies [4, 16,  20, 37] forced A. Einstein
to devise his special relativity theory and to proceed further to
creating his general relativity theory trying to explain the gravity
by means of a geometrization of space-time and matter in the
Universe. Here, we must mention that the classical Lagrangian
function $\mathcal{L}$ in (\ref {1.8}) is written by means of the
mixed combinations of terms expressed by means of both the Euclidean
``rest'' reference system variables $(r,\tau )\in \mathbb{E}^{4}$
and the arbitrarily chosen reference system variables $ (r,t)\in
\mathbb{M}^{4}.$

These problems were recently analyzed from a completely different
``no-geometry'' point of view in [6, 7, 9, 10, 20], where new
dynamical equations were derived, being free of controversy
mentioned above. Moreover, the devised approach allowed one to avoid
the introduction of the well-known Lorentz transformations of the
space-time reference systems, with respect to which the action
functional (\ref{1.8}) is invariant. From this point of view, there
are the very interesting reasonings of work [24], in which the
Galilean invariant Lagrangians possessing the intrinsic
Poincar\'{e}--Lorentz group symmetry are reanalyzed. In what
follows, we will reanalyzed the results obtained in [6, 7, 10] from
the classical Lagrangian and Hamiltonian formalisms, which will shed
a new light on the related physical backgrounds of the vacuum field
theory approach to the common study of electromagnetic and
gravitational effects.

\section{Vacuum Field Theory Electrodynamics Equations: Lagrangian
Analysis}

\subsection*{3.1. A freely moving point particle -- an\\ \hspace*{0.8cm}alternative electrodynamic
model}

Within the vacuum field theory approach to the common description of
the electromagnetism and the gravity devised in [6, 7], the main
vacuum potential field function
$\bar{W}:\mathbb{M}^{4}\mathbb{\rightarrow R} $ related to a charged
point particle $q$ satisfies, in the case of rested external charged
point objects, the dynamical equation (\ref{0.21})
%3.1
\begin{equation}
\frac{d}{dt}(-\bar{W}u)=-\nabla \bar{W},  \label{2.1}
\end{equation}
where, as above, $u:=dr/dt$ is the particle velocity with respect to some
reference system.

To analyze the dynamical equation (\ref{2.1}) from the Lagrangian point of
view, we will write the corresponding action functional as
%3.2
\begin{equation}
S:=-\underset{t_{1}}{\overset{t_{2}}{\int }}\bar{W}dt=-\underset{\tau _{1}}{%
\overset{\tau _{2}}{\int }}\bar{W}(1+\dot{r}^{2})^{1/2}\text{ }d\tau
\label{2.2}
\end{equation}
in the ``rest'' reference system $\mathcal{K}_{r}.$ Having fixed the
proper temporal parameters $\tau _{1}<\tau _{2}\in \mathbb{R}$ and
using the least action condition $\delta S=0$, one finds easily that
%3.3
\[p : =\partial \mathcal{L}/\partial \dot{r}=-\bar{W}\dot{r}(1+\dot{r}%
^{2})^{-1/2}=-\bar{W}u,\]
\begin{equation}
\dot{p} : =dp/d\tau =\partial \mathcal{L}/\partial r=-\nabla \bar{W}(1+\dot{%
r}^{2})^{1/2}.  \label{2.3}
\end{equation}
Here, owing to (\ref{2.2}), the corresponding Lagrangian function
%3.4
\begin{equation}
\mathcal{L}:=-\bar{W}(1+\dot{r}^{2})^{1/2}.  \label{2.4}
\end{equation}
Recalling now the definition of particle mass
%3.5
\begin{equation}
m:=-\bar{W}  \label{2.5}
\end{equation}
and the relations
%3.6
\begin{equation}
d\tau =dt(1-u^{2})^{1/2},\text{ }\dot{r}d\tau =udt,  \label{2.6}
\end{equation}
from (\ref{2.3}), we easily obtain exactly the dynamical equation
(\ref{2.1}). Moreover, one easily obtains that the dynamical
mass defined by expression (\ref{2.5}) is given as
\[m=m_{0}(1-u^{2})^{-1/2},\]
which coincides with result (\ref{1.3}) of the preceding section.
Thereby, based on the above-obtained results, one can formulate
the following proposition.

\noindent{\bf Proposition 3.}{\it The alternative freely moving
point particle electrodynamic model (\ref{2.1}) allows the least
action formulation (\ref{2.2}) with respect to the ``rest''
reference system variables, where the Lagrangian function is given
by expression (\ref{2.4}). Its electrodynamics is completely
equivalent to that of a classical relativistic freely moving point
particle described in Section 2.}

\subsection*{3.2. A moving charged point particle -- an\\ \hspace*{0.8cm}alternative electrodynamic
model}

Proceed now to the case where our charged point particle $q$ moves
in the space-time with velocity vector $u\in \mathbb{E}^{3}$ and
interacts with another external charged point particle, moving with
velocity vector $ u_{f}\in \mathbb{E}^{3}$ subject to some common
reference system $\mathcal{K} .$ As was shown in [6, 7], the
corresponding dynamical equation on the vacuum potential field
function $\bar{W}:\mathbb{M}^{4}\mathbb{ \rightarrow R}$ is given as
%3.7
\begin{equation}
\frac{d}{dt}[-\bar{W}(u-u_{f})]=-\nabla \bar{W}.  \label{2.7}
\end{equation}%
As the external charged particle moves in the space-time, it generates the
related magnetic field $B:=\nabla \times A,$ whose magnetic vector potential
$A:\mathbb{M}^{4}\mathbb{\rightarrow E}^{3}$ is defined, owing to the
results of \cite{BPT,BPT1,Re}, as
%3.8
\begin{equation}
qA:=\bar{W}u_{f}.  \label{2.8}
\end{equation}
Since, owing to (\ref{2.3}), the particle momentum $p=-\bar{W}u,$
Eq. (\ref{2.7}) can be equivalently rewritten as
%3.9
\begin{equation}
\frac{d}{dt}(p+qA)=-\nabla \bar{W}.  \label{2.9}
\end{equation}
To represent the dynamical equation (\ref{2.9}) within the
classical Lagrangian formalism, we start from the following action
functional naturally generalizing functional (\ref{2.2}):
%3.10
\begin{equation}
S:=-\underset{\tau _{1}}{\overset{\tau _{2}}{\int }}\bar{W}(1+|\dot{r}-\dot{%
\xi}|^{2})^{1/2}\text{ }d\tau .  \label{2.10}
\end{equation}
Here, we denoted $\dot{\xi}=u_{f}dt/d\tau $, $d\tau
=dt(1-|u-u_{f}|^{2})^{1/2}$, with regard for the relative
velocity of our charged point particle $q$ with respect to the
reference system $ \mathcal{K}_{f},$ moving with velocity vector
$u_{f}\in \mathbb{E}^{3}$ subject to the reference system
$\mathcal{K}.$ In this case, evidently, our charged point particle
$q$ moves with the velocity vector $u-u_{f}\in \mathbb{E} ^{3}$
subject to the reference system $\mathcal{K}_{f},$ and the
external charged particle is, respectively, in rest.

Compute now the least action variational condition $\delta S=0,$ taking into
account that, owing to (\ref{2.10}), the corresponding Lagrangian function
is given as
%3.11
\begin{equation}
\mathcal{L}:=-\bar{W}(1+|\dot{r}-\dot{\xi}|^{2})^{1/2}.  \label{2.11}
\end{equation}
Thereby, the total momentum of particles
%3.12
\[P :=\partial \mathcal{L}/\partial \dot{r}=-\bar{W}(\dot{r}-\dot{\xi})(1+|
\dot{r}-\dot{\xi}|^{2})^{-1/2}= \]
\[= -\bar{W}\dot{r}(1+|\dot{r}-\dot{\xi}|^{2})^{-1/2}+\bar{W}\dot{\xi}(1+|
\dot{r}-\dot{\xi}|^{2})^{-1/2}=\]
\begin{equation}
= mu+qA:=p+qA, \label{2.12}
\end{equation}
and the dynamical equation is given as
%3.13
\begin{equation}
\frac{d}{d\tau }(p+qA)=-\nabla \bar{W}(1+|\dot{r}-\dot{\xi}|^{2})^{1/2}.
\label{2.13}
\end{equation}
Since $d\tau =dt(1-\left\vert
u-u_{f}\right\vert ^{2})^{1/2}$ and $(1+|\dot{r}-\dot{\xi}
|^{2})^{1/2}=(1-|u-u_{f}|^{2})^{-1/2},$ relation
(\ref{2.13}) yields exactly the dynamical equation (\ref{2.9}). Thus, we
can formulate our result as the next proposition.

\noindent{\bf Proposition 4.} {\it The alternative classical
relativistic electrodynamic model (\ref{2.7}) allows the least
action formulation (\ref{2.10}) with respect to the ``rest''
reference system variables, where the Lagrangian function is given
by expression (\ref{2.11}).}

\subsection*{3.3. A moving charged point particle -- a dual to\\ \hspace*{0.8cm}the classical
alternative electrodynamic\\ \hspace*{0.8cm}model}

It is easy to observe that the action functional (\ref{2.10}) is written
in view of the classical Galilean transformations of reference
systems. If we now consider both the action functional (\ref{2.2}) for a charged
point particle moving with respect to the reference system $\mathcal{K}_{r}$
and its interaction with an external magnetic
field generated by the vector potential $A:$ $\mathbb{M}^{4}\rightarrow
\mathbb{E}^{3},$ it can be naturally generalized as
%3.14
\[S:=\underset{t_{1}}{\overset{t_{2}}{\int }}(-\bar{W}dt+q\langle A,dr\rangle)=\]
\begin{equation}
 =\underset{%
\tau _{1}}{\overset{\tau _{2}}{\int
}}[-\bar{W}(1+\dot{r}^{2})^{1/2} +q\langle A,\dot{r}\rangle ]d\tau
,  \label{2.14}
\end{equation}
where we accepted that $d\tau =dt(1-u^{2})^{1/2}.$

Thus, the corresponding common particle-field momentum looks as
%3.15
\[P : =\partial \mathcal{L}/\partial \dot{r}=-\bar{W}\dot{r}(1+\dot{r}
^{2})^{-1/2}+qA=\]
\begin{equation}
= mu+qA:=p+qA \label{2.15}
\end{equation}
and satisfies the equation
%3.16
\[\dot{P} : =dP/d\tau =\partial \mathcal{L}/\partial r=-\nabla \bar{W}(1+\dot{
r}^{2})^{1/2}\text{ }+q\nabla \langle A,\dot{r}\rangle =\]
\begin{equation}
 = -\nabla \bar{W}(1-u^{2})^{-1/2}+q\nabla \langle
A,u \rangle (1-u^{2})^{-1/2}, \label{2.16}
\end{equation}
where
%3.17
\begin{equation}
\mathcal{L}:= - \bar{W}(1+\dot{r}^{2})^{1/2}+q \langle A,\dot{r}
\rangle  \label{2.16a}
\end{equation}
is the corresponding Lagrangian function. Taking relation
$ d\tau =dt(1-u^{2})^{1/2}$ into account, one finds easily from
(\ref{2.16}) that
%3.18
\begin{equation}
dP/dt=-\nabla \bar{W}+q\nabla \langle A,u\rangle.  \label{2.17}
\end{equation}
By substituting (\ref{2.15}) into (\ref{2.17}) and using the well-known
\cite{LL} identity
%3.19
\begin{equation}
\nabla \langle a,b\rangle=\langle a,\nabla \rangle b+\langle
b,\nabla \rangle a+b\times (\nabla \times a)+a\times (\nabla
\times b), \label{2.18}
\end{equation}
where $a,b\in \mathbb{E}^{3}$ are arbitrary vector functions, we
obtain finally the classical expression for the Lorentz force $F$
acting on the moving charged point particle $q:$
%3.20
\begin{equation}
dp/dt:=F=qE+qu\times B.  \label{2.19}
\end{equation}
Here, by definition,
%3.21
\begin{equation}
 E:=-\nabla \bar{W}q^{-1}-\partial A/\partial t
\label{2.20}
\end{equation}
is the corresponding electric field, and
%3.22
\begin{equation}
B:=\nabla \times A  \label{2.21}
\end{equation}
is the corresponding magnetic field.

We formulate the result obtained as the next proposition.

\noindent{\bf Proposition 5.} {\it The classical relativistic
Lorentz force (\ref{2.19}) allows the least action formulation
(\ref{2.14}) with respect to the ``rest'' reference system
variables, where the Lagrangian function is given by expression
(\ref {2.16a}). Its electrodynamics described by the Lorentz force
(\ref{2.19}) is completely equivalent to the classical relativistic
moving point particle electrodynamics described by means of the
Lorentz force (\ref{1.11}) in Section 2.}

Concerning the previously obtained dynamical equation (\ref{2.13}), we can
easily observe that it can be equivalently rewritten as
%3.23
\begin{equation}
dp/dt=(-\nabla \bar{W}-qdA/dt+q\nabla \langle A,u\rangle)-q\nabla
\langle A,u\rangle. \label{2.22}
\end{equation}
The latter, owing to (\ref{2.17}) and (\ref{2.19}), takes finally
the following Lorentz-type form
%3.24
\begin{equation}
dp/dt=qE+qu\times B-q\nabla \langle A,u\rangle  \label{2.23}
\end{equation}
earlier found in [6, 7, 20].

Expressions (\ref{2.19}) and (\ref{2.23}) are equal to each other
up to the gradient term $F_{c}:=-q\nabla \langle A,u\rangle $,
which allows to reconcile the Lorentz forces acting on a charged
moving particle $q$ with respect to different reference systems.
This fact is important for our vacuum field theory approach, since
it needs to use no special geometry and makes it possible to
analyze both electromagnetic and gravitational fields
simultaneously, based on a new definition of the dynamical mass by
means of expression (\ref{2.5}).

\section{The Vacuum Field Theory Electrodynamics Equations: Hamiltonian
Analysis}

It is well known [1, 2, 8, 11, 19] that any Lagrangian theory allows
the equivalent canonical Hamiltonian representation via the
classical Legendrian transformation. As we have already formulated
our vacuum field theory of a moving charged particle $q$ in the
Lagrangian form, we proceed now to its Hamiltonian analysis making
use of the action functionals (\ref{2.2}), ( \ref{2.11}), and
(\ref{2.14}).

Let us take, first, the Lagrangian function (\ref{2.4}) and momentum
(\ref{2.3}) to define the corresponding Hamiltonian
function
%4.1
\[H : =\langle p,\dot{r}\rangle-\mathcal{L}=\]
\[= -\langle p,p\rangle\bar{W}^{-1}(1-p^{2}/\bar{W}^{2})^{-1/2}+\bar{W}(1-p^{2}/\bar{W}
^{2})^{-1/2}=\]
\[=
-p^{2}\bar{W}^{-1}(1-p^{2}/\bar{W}^{2})^{-1/2}+\bar{W}^{2}\bar{W}
^{-1}(1-p^{2}/\bar{W}^{2})^{-1/2}=\]
\begin{equation}
 = -(\bar{W}^{2}-p^{2})(\bar{W}^{2}-p^{2})^{-1/2}=-(\bar{W}^{2}-p^{2})^{1/2}.
\label{3.1}
\end{equation}
As a result, we easily obtain \cite{AM,Ar,Th,PM} \ that the
Hamiltonian function (\ref{3.1}) is a conservation law of the
dynamical field equation (\ref{2.1}). That is, for all $\tau ,t\in
\mathbb{R}$,
%4.2
\begin{equation}
dH/dt=0=dH/d\tau ,  \label{3.2}
\end{equation}
which naturally allows one to interpret it as the energy expression. Thus,
we can write that the particle energy
%4.3
\begin{equation}
\mathcal{E}=(\bar{W}^{2}-p^{2})^{1/2}.  \label{3.3}
\end{equation}
The suitable Hamiltonian equations equivalent to the vacuum field
equation ( \ref{2.1}) look as
%4.4
\[\dot{r} : =dr/d\tau =\partial H/\partial
p=p(\bar{W}^{2}-p^{2})^{-1/2}\]
\begin{equation}
\dot{p} : =dp/d\tau =-\partial H/\partial r=\bar{W}\nabla
\bar{W}(\bar{W} ^{2}-p^{2})^{-1/2}. \label{3.4}
\end{equation}

Thereby, based on the above-obtained results, one can formulate the
following proposition.

\noindent{\bf Proposition 6.} {\it The alternative freely moving
point particle electrodynamic model (\ref{2.1}) allows the canonical
Hamiltonian formulation (\ref{3.4}) with respect to the ``rest''
reference system variables, where the Hamiltonian function is given
by expression (\ref{3.1}). Its electrodynamics is completely
equivalent to the classical relativistic freely moving point
particle electrodynamics described in Section 2.}

Based now on the Lagrangian expression (\ref{2.11}), one can
construct, in the same way as above, the Hamiltonian function for the
dynamical field equation (\ref{2.9}) describing the motion of a
charged particle $q$ in an external electromagnetic field in the
canonical Hamiltonian form:
%4.5
\begin{equation}
\dot{r}:=dr/d\tau =\partial H/\partial P,\quad \dot{P}:=dP/d\tau
=-\partial H/\partial r.  \label{3.5a}
\end{equation}
Here,
%4.6
\[H : =\langle P,\dot{r}\rangle-\mathcal{L}=\]
\[=\!\langle P,\dot{\xi}\!-P\bar{W}^{-1}(1\!-P^{2}/\bar{W}^{2})^{-1/2}\rangle\!+\bar{W}[\bar{W}
^{2}(\bar{W}^{2}\!-\!P^{2})^{-1}]^{1/2}\!=\]
\[=\langle P,\dot{\xi}\rangle+P^{2}(\bar{W}^{2}-P^{2})^{-1/2}-\bar{W}^{2}(\bar{W}
^{2}-P^{2})^{-1/2}=\]
\[=-(\bar{W}^{2}-P^{2})(\bar{W}^{2}-P^{2})^{-1/2}+\langle P,\dot{\xi}\rangle=\]
\begin{equation}
=-(\bar{W}^{2}-P^{2})^{1/2}-q\langle
A,P\rangle(\bar{W}^{2}-P^{2})^{-1/2}. \label{3.5}
\end{equation}
We took into account that, owing to definitions (\ref{2.8})
and (\ref {2.12}),
%4.7
\[qA :=\bar{W}u_{f}=\bar{W}d\xi /dt=\]
\[=\bar{W}\frac{d\xi }{d\tau }\cdot \frac{d\tau
}{dt}=\bar{W}\dot{\xi} (1-\left\vert u-v\right\vert ^{2})^{1/2}=\]
\[=\bar{W}\dot{\xi}(1+|\dot{r}-\dot{\xi}|^{2})^{-1/2}=\]
\begin{equation}
 =-\bar{W}\dot{\xi}(\bar{W}^{2}-P^{2})^{1/2}\bar{W}^{-1}=-\dot{\xi}(\bar{W}
^{2}-P^{2})^{1/2},  \label{3.6}
\end{equation}
or
%4.8
\begin{equation}
\dot{\xi}=-qA(\bar{W}^{2}-P^{2})^{-1/2},  \label{3.7}
\end{equation}
where $A:\mathbb{M}^{4}\mathbb{\rightarrow R}^{3}$ is the related magnetic
vector potential generated by the moving external charged particle.

Thereby, we can state that the Hamiltonian function (\ref{3.5}) satisfies the
energy conservation conditions%
%4.9
\begin{equation}
dH/dt=0=dH/d\tau  \label{3.8}
\end{equation}
for all $\tau ,t\in \mathbb{R}.$ That is, the suitable energy
expression
%4.10
\begin{equation}
\mathcal{E}=(\bar{W}^{2}-P^{2})^{1/2}+q\langle
A,P\rangle(\bar{W}^{2}-P^{2})^{-1/2} \label{3.9}
\end{equation}
holds. Result (\ref{3.9}) essentially differs from that
obtained in [13] which makes use of the well-known Einsteinian
Lagrangian for a moving charged point particle $q$ in an external
electromagnetic field. Thereby, our result can be formulated as
follows.

\noindent{\bf Proposition 7.} {\it The alternative classical
relativistic electrodynamic model (\ref{2.7}) allows the
Hamiltonian formulation (\ref{3.5a}) with respect to the ``rest''
reference system variables, where the Hamiltonian function is
given by expression (\ref{3.5}).}

To make this difference more clear, we will analyze below the
Lorentz force (\ref{2.19}) from the Hamiltonian point of view
based on the Lagrangian function (\ref{2.16a}). Thus, we obtain
that the corresponding Hamiltonian
function%
%4.11
\[H :=\langle P,\dot{r}\rangle-\mathcal{L}=\langle
P,\dot{r}\rangle+\bar{W}(1+\dot{r}^{2})^{1/2}-q\langle A,
\dot{r}\rangle=\]
\[=\langle P-qA,\dot{r}\rangle+\bar{W}(1+\dot{r}^{2})^{1/2}=\]
\[=-\langle p,p\rangle\bar{W}^{-1}(1-p^{2}/\bar{W}^{2})^{-1/2}+\bar{W}(1-p^{2}/\bar{W}
^{2})^{-1/2}=\]
\begin{equation}
=-(\bar{W}^{2}-p^{2})(\bar{W}^{2}-p^{2})^{-1/2}=-(\bar{W}^{2}-p^{2})^{1/2}.
\label{3.10}
\end{equation}
Since $p=P-qA,$ expression (\ref{3.10}) takes the final ``\textit{no
interaction}''  [13, 42, 46, 47] form
%4.12
\begin{equation}
H=-[\bar{W}^{2}-(P-qA)^{2}]^{1/2},  \label{3.11}
\end{equation}
being conservative with respect to the evolution equations
(\ref{2.15}) and ( \ref{2.16}), i.e.,
%4.13
\begin{equation}
dH/dt=0=dH/d\tau  \label{3.11a}
\end{equation}
for all $\tau ,t\in \mathbb{R}.$ The latter are simultaneously
equivalent to the following Hamiltonian system:
%4.14
\[\dot{r} =\partial H/\partial
P=(P-qA)[\bar{W}^{2}-(P-qA)^{2}]^{-1/2},\]
 \[\dot{P} =-\partial H/\partial r=(\bar{W}\nabla \bar{W}-\nabla \langle qA,(P-qA)\rangle)\times\]
\begin{equation}
\times[ \bar{W}^{2}-(P-qA)^{2}]^{-1/2}, \label{3.12}
\end{equation}
that can be easily checked by direct calculations. Really, the
first equation
%4.15
\[\dot{r}
=(P-qA)[\bar{W}^{2}-(P-qA)^{2}]^{-1/2}=p(\bar{W}^{2}-p^{2})^{-1/2}=\]
\begin{equation}
=mu(\bar{W}^{2}-p^{2})^{-1/2}=-\bar{W}u(\bar{W}
^{2}-p^{2})^{-1/2}=u(1-u^{2})^{-1/2} \label{3.13}
\end{equation}
holds, owing to the condition $d\tau =dt(1-u^{2})^{1/2}$ and
definitions $ p:=mu$ and $m=-\bar{W}$ postulated from the very
beginning. Similarly, we obtain
%4.16
\[\dot{P} =-\nabla \bar{W}(1-p^{2}/\bar{W}^{2})^{-1/2}+\nabla \langle qA,u\rangle(1-p^{2}/
\bar{W}^{2})^{-1/2}=\]
\begin{equation}
=-\nabla \bar{W}(1-u^{2})^{-1/2}+\nabla \langle
qA,u\rangle(1-u^{2})^{-1/2}, \label{3.14}
\end{equation}
exactly coinciding with Eq. (\ref{2.17}) subject to the
evolution parameter $t\in \mathbb{R}$. We now formulate
our result as the next proposition.

\noindent{\bf Proposition.} {\it Model (\ref{2.19}) dual to the
classical relativistic electrodynamics model allows the Hamiltonian
formulation (\ref{3.12}) with respect to the ``rest'' reference
system variables, where the Hamiltonian function is given by
expression (\ref{3.11}).}

\section{The Quantization of Electrodynamics Models within the Vacuum Field
Theory no-Geometry Approach}

\subsection*{5.1. Statement of the problem}

In our recent works [6, 7], there was devised a new regular
no-geometry approach to deriving, from the first principles, the
electrodynamics of a moving charged point particle $q$ in an
external electromagnetic field. This approach has, in part, to
reconcile the existing mass-energy controversy [16] within the
classical relativistic electrodynamics. Based on the vacuum field
theory approach proposed in [6, 7, 20], we have reanalyzed this
problem in the sections above both from the Lagrangian and
Hamiltonian points of view, having derived crucial expressions for
the corresponding energy functions and Lorentz-type forces acting on
moving charge point particle $q$.

Since all of our electrodynamics models were represented here in the
canonical Hamiltonian form, they are suitable for applying the
Dirac-type quantization procedure to them [3, 12, 15] and for
obtaining the related Schr\"{o}dinger-type evolution equations. The
following section is devoted namely to this problem.

\subsection*{5.2. Free point particle electrodynamics model\\ \hspace*{0.8cm}and its quantization}

The charged point particle electrodynamics models discussed in
Sections 2 and 3 in detail were also considered in [7] from the
dynamical point of view, where an attempt to apply the quantization
Dirac-type procedure to the corresponding conserved energy
expressions was done. Nevertheless, within the canonical point of
view, the true quantization procedure should be based on the
suitable canonical Hamiltonian formulation of the models which looks
in the case under consideration as (\ref{3.4}), (\ref{3.5a}), and
(\ref{3.12}).

In particular, consider the free charged point particle electrodynamics
model (\ref{3.4}) governed by the Hamiltonian equations%
%5.1
\[dr/d\tau :=\partial H/\partial
p=-p(\bar{W}^{2}-p^{2})^{-1/2},\]
\begin{equation}
dp/d\tau :=-\partial H/\partial r=-\bar{W}\nabla \bar{W}(\bar{W}
^{2}-p^{2})^{-1/2}, \label{5.1}
\end{equation}
where we denoted, as before, the corresponding vacuum field potential
characterizing a medium field structure by $\bar{W}:\mathbb{M}^{4}\rightarrow
\mathbb{R}$, the standard canonical
coordinate-momentum variables by $(r,p)\in
\mathbb{E}^{3}\times \mathbb{E}^{3}$, and the proper
``rest'' reference system $\mathcal{K}_{r}$ time parameter
related to our moving particle by $\tau \in \mathbb{R}$. The notation $H:\mathbb{E}^{3}\times
\mathbb{E}^{3}\rightarrow \mathbb{R}$ stands for the Hamiltonian function
%5.2
\begin{equation}
H:=-(\bar{W}^{2}-p^{2})^{1/2}  \label{5.2}
\end{equation}
expressed here and throughout further in the light speed units.
The ``rest'' reference system $\mathcal{K}_{r}$ parametrized by the
variables $(r,\tau )\in \mathbb{E}^{4}$ is related to any other
reference system $\mathcal{K}$, subject to which our charged point
particle $q$ moves with a velocity vector $ u\in \mathbb{E}^{3}$,
and which is parametrized by the variables $(r,t)\in \mathbb{M}^{4}$
via the Euclidean infinitesimal relation
%5.3
\begin{equation}
dt^{2}=d\tau ^{2}+dr^{2}  \label{5.3}
\end{equation}
which is equivalent to the Minkowskian infinitesimal relation
%5.4
\begin{equation}
d\tau ^{2}=dt^{2}-dr^{2}.  \label{5.4}
\end{equation}
The Hamiltonian function (\ref{5.2}) satisfies, evidently, the energy
conservation conditions
%5.5
\begin{equation}
dH/dt=0=dH/d\tau  \label{5.5}
\end{equation}
for all $t,\tau \in \mathbb{R}$. This means that the suitable
energy value
%5.6
\begin{equation}
\mathcal{E}=(\bar{W}^{2}-p^{2})^{1/2}  \label{5.6}
\end{equation}
can be treated by means of the Dirac-type quantization scheme [12]
to obtain, as $\hbar \rightarrow 0 $, (or the light speed
$c\rightarrow \infty $ ) the governing Schr\"{o}dinger-type
dynamical equation. To do this, similarly to [6, 7], we need to make
canonical operator replacements $\mathcal{E}\rightarrow
\hat{\mathcal{E}}:=-\frac{\hbar }{i} \frac{\partial }{\partial \tau
},$ \ $p\rightarrow \hat{p}:=\frac{\hbar }{i} \nabla $, as $\hbar
\rightarrow 0$, in the energy-determining expression
%5.7
\begin{equation}
\mathcal{E}^{2}:=(\hat{\mathcal{E}}\psi ,\hat{\mathcal{E}}\psi
)=(\psi ,\hat{ \mathcal{E}}^{2}\psi )=(\psi
,\hat{H}^{+}\hat{H}\psi ).  \label{5.7}
\end{equation}
Here, by definition, owing to (\ref{5.6}),
%5.8
\begin{equation}
\hat{\mathcal{E}}^{2}=\bar{W}^{2}-\hat{p}^{2}=\hat{H}^{+}\hat{H}  \label{5.8}
\end{equation}
is a suitable operator factorization in the Hilbert space
$\mathcal{H} :=L_{2}(\mathbb{R}^{3};\mathbb{C})$, and $\psi \in
\mathcal{H}$ is the corresponding normalized quantum vector state.
Since the elementary identity
%5.9
\begin{equation}
\bar{W}^{2}-\hat{p}^{2}=\bar{W}(1-\bar{W}^{-1}\hat{p}^{2}\bar{W}
^{-1})^{1/2}(1-\bar{W}^{-1}\hat{p}^{2}\bar{W}^{-1})^{1/2}\bar{W}
\label{5.9}
\end{equation}
holds, we can set, by definition, following (\ref{5.8}) and (\ref{5.9}),
the operator
%5.10
\begin{equation}
\hat{H}:=(1-\bar{W}^{-1}\hat{p}^{2}\bar{W}^{-1})^{1/2}\bar{W}.  \label{5.10}
\end{equation}
Having calculated the operator expression (\ref{5.10}) as $\hbar
\rightarrow 0$ up to operator accuracy $O(\hbar ^{4})$, we obtain
easily
%5.11
\begin{equation}
\hat{H}=\frac{\hat{p}^{2}}{2m(u)}+\bar{W}:=-\frac{\hbar ^{2}}{2m(u)}\nabla
^{2}+\bar{W}  \label{5.11}
\end{equation}
with regard for the dynamical mass definition $m(u):=-\bar{W}$ (in
the light speed units). Thereby, based now on (\ref{5.7}) and
(\ref{5.11}), we obtain, up to operator accuracy $O(\hbar ^{4})$,
the Schr\"{o}dinger-type evolution equation
%5.12
\begin{equation}
i\hbar \frac{\partial \psi }{\partial \tau
}:=\hat{\mathcal{E}}\psi =\hat{H} \psi =-\frac{\hbar
^{2}}{2m(u)}\nabla ^{2}\psi +\bar{W}\psi  \label{5.12}
\end{equation}
with respect to the ``rest'' reference system $\mathcal{K}_{r}$
evolution parameter $\tau \in \mathbb{R}$. Concerning the related
evolution parameter $ t\in \mathbb{R}$ parametrizing a reference
system $\mathcal{K}$, Eq. (\ref{5.12}) takes the
form
%5.13
\begin{equation}
i\hbar \frac{\partial \psi }{\partial t}=-\frac{\hbar
^{2}m_{0}}{2m(u)^{2}} \nabla ^{2}\psi -m_{0}\psi .  \label{5.13}
\end{equation}
Here we took into account that, owing to (\ref{5.6}), the classical mass
relation
%5.14
\begin{equation}
m(u)=m_{0}(1-u^{2})^{-1/2}  \label{5.14}
\end{equation}
holds, where $m_{0}\in \mathbb{R}_{+}$ is the corresponding rest
mass of our point particle $q$.

As $ \hbar /c\rightarrow 0$, the obtained linear Schr\"{o}dinger
equation (\ref{5.13}) coincides really with the well-known equation
[13, 12, 5] from classical quantum mechanics.

\subsection*{5.3. Classical charged point particle\\ \hspace*{0.8cm}electrodynamics model and its\\ \hspace*{0.8cm}quantization}

We start here from the first vacuum field theory reformulation of the
classical charged point particle electrodynamics considered in Section 3
and based on the conserved Hamiltonian function (\ref{3.11})
%5.15
\begin{equation}
H:=-[\bar{W}^{2}-(P-qA)^{2}]^{1/2}.  \label{6.1}
\end{equation}
Here, $q\in $ $\mathbb{R}$ is the particle charge, $(\bar{W},A)\in
\mathbb{R\times E}^{3}$ is the corresponding electromagnetic field
potentials, and $P\in \mathbb{E}^{3}$ is the common particle-field momentum
defined as
%5.16
\begin{equation}
P:=p+qA,\quad p:=mu  \label{6.2}
\end{equation}
and satisfying the well-known classical Lorentz force equation.
Here, $m:=- \bar{W}$ is the observable dynamical mass related to
our charged particle, and $u\in \mathbb{E}^{3}$ is its velocity vector
with respect to a chosen reference system $\mathcal{K}$, being all
expressed here, as before, in the light speed units.

As our electrodynamics based on (\ref{6.1}) is canonically Hamiltonian,
the Dirac-type quantization scheme
%5.17
\begin{equation}
P\rightarrow \hat{P}:=\frac{\hbar }{i}\nabla ,\quad \mathcal{E}
\rightarrow \hat{\mathcal{E}}:=-\frac{\hbar }{i}\frac{\partial
}{\partial \tau }  \label{6.3}
\end{equation}
should be applied to the suitable energy expression
%5.18
\begin{equation}
\mathcal{E}:=[\bar{W}^{2}-(P-qA)^{2}]^{1/2}  \label{6.4}
\end{equation}
following from the conservation conditions
%5.19
\begin{equation}
dH/dt=0=dH/d\tau  \label{6.5}
\end{equation}
satisfied for all $\tau ,t\in \mathbb{R}$.

Passing now the same way as above, we can factorize the operator
$\hat{E}^{2}$ as follows:
\[\bar{W}^{2}-(\hat{P}-qA)^{2}=\bar{W}[1-\bar{W}^{-1}(\hat{P}-qA)^{2}\bar{W}
^{-1}]^{1/2}\times\]
\[\times \lbrack 1-\bar{W}^{-1}(\hat{P}-qA)^{2}\bar{W}^{-1}]^{1/2}\bar{W}:=
\hat{H}^{+}\hat{H}.\] Here, by definition (here as $\hbar
/c\rightarrow 0,$ $\hbar c=$~const),
%5.20
\begin{equation}
\hat{H}:=\frac{1}{2m(u)}(\frac{\hbar }{i}\nabla -qA)^{2}+\bar{W}  \label{6.7}
\end{equation}
up to operator accuracy $O(\hbar ^{4})$. Thereby, the related
Schr\"{o}dinger-type evolution equation in the Hilbert space
$\mathcal{H}=L_{2}( \mathbb{R}^{3};\mathbb{C})$ looks as
%5.21
\begin{equation}
i\hbar \frac{\partial \psi }{\partial \tau
}:=\hat{\mathcal{E}}\psi =\hat{H} \psi
=\frac{1}{2m(u)}(\frac{\hbar }{i}\nabla -qA)^{2}\psi +\bar{W}\psi
\label{6.8}
\end{equation}
with respect to the rest reference system $\mathcal{K}_{r}$
evolution parameter $\tau \in \mathbb{R}$. The corresponding
Schr\"{o}dinger-type evolution equation with respect to the
evolution parameter $t\in \mathbb{R}$ looks, respectively, as
%5.22
\begin{equation}
i\hbar \frac{\partial \psi }{\partial t}=\frac{m_{0}}{2m(u)^{2}}(\frac{\hbar
}{i}\nabla -qA)^{2}\psi -m_{0}\psi .  \label{6.9}
\end{equation}
The Schr\"{o}dinger-type evolution equation (\ref{6.8}) (as $\hbar
/c\rightarrow 0,$ $\hbar c=$~const) completely coincides [14, 12]
with that well known from the classical quantum mechanics.

\subsection*{5.4. Modified charged point particle\\ \hspace*{0.8cm}electrodynamics model and its\\ \hspace*{0.8cm}quantization}

Consider now, within the canonical point of view, the true quantization
procedure of the electrodynamics model, which looks as (\ref{2.13}) and whose
Hamiltonian function (\ref{3.5}) is
%5.23
\begin{equation}
H:=-(\bar{W}^{2}-P^{2})^{1/2}-q\langle
A,P\rangle(\bar{W}^{2}-P^{2})^{-1/2}. \label{7.1}
\end{equation}
This means that the suitable energy function
%5.24
\begin{equation}
\mathcal{E}:=(\bar{W}^{2}-P^{2})^{1/2}+q\langle
A,P\rangle(\bar{W}^{2}-P^{2})^{-1/2}, \label{7.2}
\end{equation}
where, as before,
%5.25
\begin{equation}
P:=p+qA,\quad p:=mu,\quad m:=-\dot{W},  \label{7.3}
\end{equation}
is a conserved quantity for (\ref{2.13}), which we will
canonically quantize via the Dirac procedure (\ref{6.3}). To make
this, let us consider the quantum condition
%5.26
\begin{equation}
\mathcal{E}^{2}:=(\hat{\mathcal{E}}\psi ,\hat{\mathcal{E}}\psi
)=(\psi ,\hat{ \mathcal{E}}^{2}\psi ),\quad (\psi ,\psi ):=1,
\label{7.4}
\end{equation}
where, by definition, $\hat{\mathcal{E}}:=-\frac{\hbar
}{i}\frac{\partial }{
\partial t}$, and $\psi \in \mathcal{H}=L_{2}(\mathbb{R}^{3};\mathbb{C})$ is
a suitable normalized quantum state vector. Making now use of the energy
function (\ref{7.2}), one can easily obtain
%5.27
\[\mathcal{E}^{2}=\bar{W}^{2}-(P-qA)^{2}+\]
\begin{equation}
+q^{2}[\langle A,A\rangle+\langle A,P\rangle(\bar{W}
^{2}-P^{2})^{-1}\langle P,A\rangle]  \label{7.5}
\end{equation}
which transforms upon the canonical Dirac-type quantization $P\rightarrow
\hat{P}:= \frac{\hbar }{i}\nabla $ into the symmetrized
operator expression
%5.28
\[\hat{\mathcal{E}}^{2}=\bar{W}^{2}-(\hat{P}-qA)^{2}+\]
\begin{equation}
+q^{2}\langle A,A\rangle+q^{2}\langle A,\hat{P}
\rangle(\bar{W}^{2}-\hat{P}^{2})^{-1}\langle\hat{P},A\rangle.
\label{7.6}
\end{equation}
Having factorized operator (\ref{7.6})\ in the form
$\hat{\mathcal{E}}^{2}= \hat{H}^{+}\hat{H},$ we obtain, up to
operator accuracy $O(\hbar ^{4})$ (as $\hbar /c\rightarrow 0$,
$\hbar c=$~const),
%15.29
\[\hat{H} :=\frac{1}{2m(u)}(\frac{\hbar }{i}\nabla -qA)^{2}-\]
\begin{equation}
-\frac{q^{2}}{2m(u)} \langle
A,A\rangle-\frac{q^{2}}{2m^{3}(u)}\langle A,\frac{\hbar }{i}
\nabla \rangle\langle\frac{\hbar }{i}\nabla ,A\rangle, \label{7.7}
\end{equation}
where we put, as before, $m(u)=-\bar{W}$ in the light speed units. Thus,
owing to (\ref{7.4}) and (\ref{7.7}), the resulting Schr\"{o}dinger
evolution equation takes the form
%5.30
\[i\hbar \frac{\partial \psi }{\partial \tau } :=\hat{H}\psi =\frac{1}{2m(u)}
(\frac{\hbar }{i}\nabla -qA)^{2}\psi -\]
\begin{equation}
-\frac{q^{2}}{2m(u)} \langle A,A\rangle\psi
-\frac{q^{2}}{2m^{3}(u)}\langle A,\frac{\hbar }{i} \nabla
\rangle\langle\frac{\hbar }{i}\nabla ,A\rangle\psi \label{7.8}
\end{equation}
with respect to the ``rest'' reference system proper evolution
parameter $\tau \in \mathbb{R}$. Similarly, one obtains also the
related Schr\"{o}dinger-type evolution equation with respect to the
time parameter $t\in \mathbb{R}$, on which we will not stop here.
Result (\ref{7.8}) essentially differs from the corresponding
classical Schr\"{o}dinger evolution equation (\ref{6.8}), which
forces us, thereby, to reanalyze more thoroughly the main physically
motivated principles put into the backgrounds of the classical
electrodynamic models described by the Hamiltonian functions
(\ref{6.1}) and (\ref{7.1}) giving rise to different Lorentz-type
force expressions. We plan to do this analysis in a next work under
preparation in detail.

\section{Conclusion}

Based on the results obtained,  we can claim that all of the
electrodynamical field equations discussed above are equivalent to
canonical Hamiltonian flows with respect to the corresponding proper
``rest'' reference systems parametrized by suitable time parameters
$\tau \in \mathbb{R}$. Owing to the passing to the laboratory
reference system $\mathcal{K}$ parametrized with the time parameter
$t\in \mathbb{R}$, the related Hamiltonian structures appear to be
naturally lost, giving rise to a new interpretation of the real
charged particle motion as such one having the absolute sense only
with respect to the proper ``rest'' reference system and being
completely relative with respect to all other reference systems.
Concerning the Hamiltonian expressions (\ref{3.1}), (\ref{3.5}), and
(\ref {3.11}) obtained above, one observes that all of them depend
strongly on the vacuum potential field function
$\bar{W}:\mathbb{M}^{4}\mathbb{\rightarrow R} $, thereby dissolving
the mass problem of the classical energy expression, before pointed
out \cite{Br} by L. Brillouin. We mention here that, subject to the
canonical Dirac-type quantization procedure, it can be applied only
to the corresponding dynamical field systems considered with respect
to their proper ``rest'' reference systems.\looseness=1

\noindent{\bf Remark 9.} {\it Some comments can be also made
concerning the classical relativity principle. Namely, we have
obtained our results completely without using the Lorentz
transformations of reference systems but only the natural notion of
the ``rest'' reference system and its suitable parametrization with
respect to any other moving reference systems. It looks reasonable,
since the true state changes of a moving charged particle $q$ are
exactly realized in reality only with respect to its proper ``rest''
reference system. Thereby, the only question still here left open is
that about the physical justification of the corresponding relation
between the time parameters of the moving and ``rest'' reference
systems.}\looseness=1

This relation, being accepted throughout this work, looks as
%6.1
\begin{equation}
d\tau =dt(1-u^{2})^{1/2},  \label{4.1}
\end{equation}
where $u:=dr/dt\in \mathbb{E}^{3}$ is the velocity vector, with
which the ``rest'' reference system $\mathcal{K}_{r}$ moves with
respect to another arbitrarily chosen reference system
$\mathcal{K}$. Expression (\ref{4.1}) means, in particular, that the
equality
%6.2
\begin{equation}
dt^{2}-dr^{2}=d\tau ^{2}  \label{4.2}
\end{equation}
holds, and it exactly coincides with the classical infinitesimal
Lorentz invariant. Its appearance is, evidently, not casual here,
since all our dynamical vacuum field equations were successively
derived [6, 7] from the governing equations on the vacuum potential
field function $W:\mathbb{M}^{4} \mathbb{\rightarrow R}$ in the form
%6.3
\[\partial ^{2}W/\partial t^{2}-\nabla ^{2}W=\rho ,\]
\begin{equation}
\partial W/\partial
t+\nabla (vW)=0,\quad \partial \rho /\partial t+\nabla (v\rho )=0,
\label{4.3}
\end{equation}
being {\it a priori} Lorentz invariant, where we denoted the charge density by $\rho
\in $ $ \mathbb{R}$ and the
suitable local velocity of vacuum field potential changes by $v:=dr/dt$.
Thereby, the dynamical infinitesimal Lorentz invariant (\ref{4.2})
reflects this intrinsic structure of Eqs. (\ref{4.3}). Being
rewritten in the nonstandard Euclidean form
%6.4
\begin{equation}
dt^{2}=d\tau ^{2}+dr^{2},  \label{4.4}
\end{equation}
it gives rise to a completely other time relation between
reference systems $\mathcal{K}$ and $\mathcal{K}_{r}:$
%6.5
\begin{equation}
dt=d\tau (1+\dot{r}^{2})^{1/2},  \label{4.5}
\end{equation}
where we denoted, as earlier, the related particle velocity with
respect to the ``rest'' reference system by $\dot{r}:=dr/d\tau $.
Thus, we observe that all our Lagrangian analysis completed in
Section 2 is based on the corresponding functional expressions
written in these ``Euclidean'' space-time coordinates and with
respect to which the least action principle was applied. So, we see
that there exist two alternatives -- the first is to apply the least
action principle to the corresponding Lagrangian functions expressed
in the Minkowski-type space-time variables with respect to an
arbitrary chosen laboratory reference system $\mathcal{K}$, and the
second is to apply the least action principle to the corresponding
Lagrangian functions expressed in the space-time Euclid-type
variables with respect to the ``rest'' reference system
$\mathcal{K}_{r}$.

As a slightly amusing but exciting inference, following from our
analysis in this work, is the fact that all of the classical special
relativity results related to the electrodynamics of charged point
particles can be obtained one-to-one, by making use of our new
definitions of the dynamical particle mass and the least action
principle with respect to the associated Euclid-type space-time
variables parametrizing the ``rest'' reference system.

An additional remark is needed concerning the quantization procedure
of the proposed electrodynamics models. If the dynamical vacuum
field equations are expressed in the canonical Hamiltonian form,
only technical problems are left to quantize them and to obtain the
corresponding Schr\"{o}dinger-type evolution equations in suitable
Hilbert spaces of quantum states. There exists still another
important inference from the approach devised in this work. It
consists in the complete lost of the essence of the well-known
Einsteinian equivalence principle [4, 5, 13,  37, 42] becoming
superfluous for our vacuum field theory of electromagnetism and
gravity.

Based on the canonical Hamiltonian formalism devised in this work,
concerning the alternative charged point particle electrodynamics
models, we succeeded in treating their Dirac-type quantization. The
obtained results were compared with classical ones, but the
physically motivated choice of a true model is left for the future
studies. Another important aspect of the developed vacuum field
theory no-geometry approach to combining the electrodynamics with
the gravity consists in singling out the decisive role of the
related ``rest'' reference system $\mathcal{K}_{r}$. Namely, with
respect to the ``rest'' reference system evolution parameter $\tau
\in \mathbb{R}$, all of our electrodynamics models allow both the
Lagrangian and Hamiltonian formulations suitable for the canonical
quantization. The physical nature of this fact remains, by now, not
enough understood. There is, by now [13, 30--32, 37, 40--42], no
physically reasonable explanation of this decisive role of the
``rest'' reference system, except for the very interesting
reasonings by R. Feynman who argued in [5] that the relativistic
expression for the classical Lorentz force (\ref{1.11}) has physical
sense only with respect to the ``rest'' reference system variables
$(r,\tau )\in \mathbb{E}^{4}$. In the sequel of our work, we plan to
analyze the quantization scheme in more details and make a step
toward the vacuum quantum field theory of infinite-many-particle
systems.\looseness=1

\vskip3mm The authors are cordially thankful to the Abdus Salam
International Centre for Theoretical Physics in Trieste, Italy, for
the hospitality during their research 2007--2008 scholarships. A.P.
is, especially, grateful to his friends and colleagues Profs. P.I.
Holod (UKMA, Kyiv), J.M.~Stakhira (Lviv, NUL), U. Taneri (Cyprus,
EMU), Z. Peradzy{\'{n}}ski (Warsaw, UW) and J. S\l awianowski
(Warsaw, IPPT) for fruitful discussions, useful comments, and
remarks. The authors are also appreciated to Profs. T.L. Gill, W.W.
Zachary, and J. Lindsey for some related references, comments, and
sending their very interesting preprint \cite{GZ} before its
publication.  The last but not least thanks go to Academician Prof.
A.A.~Logunov for his interest to the work, to Referees for some
instrumental  suggestions, as well as to Mrs. Dilys Grilli (Trieste,
Publications office, ICTP) and Natalia K. Prykarpatska for the
professional help in preparing the manuscript for publication.


\begin{thebibliography}{99}
\bibitem{Ar} V.I. Arnold, {\it Mathematical Methods of Classical Mechanics}
(Springer, Berlin, 1997).

\bibitem{AM} R. Abraham and J. Marsden, {\it Foundations of Mechanics}
(Benjamin/Cummings, New York, 1978).

\bibitem{BS} N.N. Bogolyubov and D.V. Shirkov, {\it Introduction to the Theory of
Quantized Fields} (Nauka, Moscow, 1984) (in Russian).

\bibitem{Fe} R. Feynman, {\it Lectures on Gravitation} (California
Inst. of Technology, 1971).

\bibitem{Fe1} R. Feynman, R. Leighton, and M. Sands, {\it The Feynman Lectures on
Physics. Electrodynamics}, v. 2 (Addison-Wesley, New York, 1964).

\bibitem{BPT} A.K. Prykarpatsky, N.N. Bogolubov, jr., and U.
Taneri, arXiv lanl: 0807.3691v.8 [gr-gc] (24.08.2008).

\bibitem{BPT1} A.K. Prykarpatsky, N.N. Bogolubov, jr., and U. Taneri, {\it The
field structure of vacuum, Maxwell equations and relativity theory
aspects} (Preprint ICTP, Trieste, IC/2008/051)
(http://publications.ictp.it).

\bibitem{PM} A. Prykarpatsky and I. Mykytiuk, {\it Algebraic Integrability of
Nonlinear Dynamical Systems on Manifolds: Classical and Quantum
Aspects} (Kluwer, Dordrecht, 1998).

\bibitem{BP1} N.N. Bogolubov, jr., and A.K. Prykarpatsky, {\it The vacuum structure,
special relativity and quantum mechanics revisited: a field theory
no-geometry approach within the Lagrangian and Hamiltonian
formalisms}. Part 2. Preprint ICTP, Trieste, IC/2008/091 (available
at: http://publications.ictp.it IC/2008/091).

\bibitem{BPT2} A.K. Prykarpatsky and N.N. Bogolubov, jr., arXiv:0810.3755v1 [gr-qc] 21
Oct (2008).

\bibitem{Th} W. Thirring, {\it Classical Mathematical Physics} (Springer,
Berlin, 1992).

\bibitem{Di} P.A.M. Dirac, {\it The Principles of Quantum Mechanics}
(Clarendon Press, Oxford, 1935).

\bibitem{LL} L.D Landau and E.M. Lifshits, The Classical Theory of
Fields (Pergamon Press, Oxford, 1983).

\bibitem{LL1} L.D. Landau and E.M. Lifshitz, {\it Quantum Mechanics}
(Pergamon Press, New York, 1980).

\bibitem{BS1} N.N. Bogolubov and D.V. Shirkov, {\it Introduction to the Theory of
Quantized Fields} (Interscience, New York, 1959).

\bibitem{Br} L. Brillouin,  {\it Relativity Reexamined} (Academic Press, New
York, 1970).

\bibitem{Fa} L.D. Faddeev, Uspekhi Fiz. Nauk {\bf 136}, 435 (1982).

\bibitem{MC} J. Marsden  and A. Chorin, {\it Mathematical Foundations of the
Mechanics of Liquid} (Springer, New York, 1993).

\bibitem{HK} P.I. Holod and A.I. Klimyk, {\it Mathematical Foundations of Symmetry
Theory} (Naukova Dumka, Kyiv, 1992) (in Ukrainian).

\bibitem{Re} O. Repchenko, {\it Field Physics} (Galeria, Moscow,
2005).

\bibitem{BD} C.H. Brans and R.H. Dicke, Phys. Rev. {\bf 124}, 925 (1961).

\bibitem{DJ} S. Deser and R. Jackiw, arXiv:hep-th/9206094 (1992).

\bibitem{DuJ} G. Dunner and R. Jackiw, arXiv:hep-th/92004057 (1992).

\bibitem{RP} R. Jackiw and A.P. Polychronakos, arXiv:hep-th/ 9809123 (1998).

\bibitem{BPZZ} B.M. Barbashov, V.N. Pervushin, A.F. Zakharov, and V.A.
Zinchuk, arXiv:hep-th/0606054 (2006).

\bibitem{Ba} B.M. Barbashov, arXiv:hepth/0111164 (2001).

\bibitem{Ja} R. Jackiw, arXiv:hep-th/0709.2348 (2007).

\bibitem{B-W} I.E. Bulyzhenkov-Widicker, Int. J. of Theor.
Phys. {\bf 47}, 1261 (2008).

\bibitem{Bu} I.E. Bulyzhenkov, arXiv:math-ph/0603039 (2008).

\bibitem{GZL} T.L. Gill, W.W. Zachary, and J. Lindsey, Found. of Physics {\bf 31}, 1299 (2001).

\bibitem{GZ} T.L. Gill and W.W. Zachary, {\it Two mathematically equivalent versions
of Maxwell equations} (Preprint, University of Maryland, 2008).

\bibitem{Da} T. Damour, Ann. Phys. (Leipzig) {\bf 17}, 619  (2008).\vskip2mm

\bibitem{Wi} F. Wilchek, Ann. Henry Poincar\'{e} {\bf 4},
211 (2003).\vskip2mm

\bibitem{B-B} I. Bialynicky-Birula, Phys Rev. {\bf 155}, 1414 (1967); Phys Rev. {\bf 166}, 1505 (1968).\vskip2.8mm

\bibitem{Sl} J.J. S{\l }awianowski, {\it Geometry of Phase Spaces} (Wiley,
New York, 1991).\vskip2mm

\bibitem{Kle} H. Kleinert, {\it Path Integrals} (World Scientific, Singapore,
1995).\vskip2mm

\bibitem{Kl} I.A. Klymyshyn, {\it Relativistic Astronomy} (Naukova Dumka, Kyiv,
1980) (in Ukrainian).\vskip2mm

\bibitem{Lo3} A.A. Logunov and M.A. Mestvirishvili, {\it Relativistic Theory of
Gravitation} (Nauka, Moscow, 1989) (in Russian).\vskip2mm

\bibitem{Lo2} A.A. Logunov, {\it The Theory of Gravity} (Nauka,
Moscow, 2000) (in Russian).\vskip2mm

\bibitem{Lo1} A.A. Logunov, {\it Relativistic Theory of Gravitation}
(Nauka, Moscow, 2006) (in Russian).\vskip2mm

\bibitem{Lo} A.A. Logunov, {\it Lectures on Relativity Theory and Gravitation}
(Nauka, Moscow, 1987) (in Russian).\vskip2mm

\bibitem{Pa} W. Pauli, {\it Theory of Relativity} (Dover, New York,
1981).\vskip2mm

\bibitem{We} R. Weinstock, Am. J. Phys.
{\bf 33}, 640 (1965).\vskip2mm

\bibitem{Me1} N.D. Mermin, Am. J. Phys.
{\bf 52}, 119 (1984).\vskip2mm

\bibitem{Me} N.D. Mermin, \textit{It's About Time: Understanding Einstein's
Relativity} (Princeton Univ. Press, Princeton, 2005).\vskip2mm

\bibitem{Ku} B.A. Kupershmidt, Diff. Geom. Appl. {\bf 2}, 275 (1992).\vskip2mm

\bibitem{PSP} Y.A. Prykarpatsky, A.M. Samoilenko, and A.K. Pry\-kar\-patsky, Opuscula
Math. {\bf 25}, 287 (2005).\vskip2mm

\bibitem{Gr1} B. Green, {\it The Fabric of the Cosmos} (Vintage, New
York, 2004). \vskip2mm

\bibitem{Ne} R.P. Newman, Comm. Mathem. Phys. {\bf 123}, 17 (1989).%\vskip3mm

\begin{flushright}
{\footnotesize Received 23.02.09}
\end{flushright}
\end{thebibliography}
\end{document}